\documentclass[a4paper,11pt]{article}
\usepackage{pos}
\usepackage{tikz}
\usetikzlibrary{positioning}

\title{Bootstrapping two-loop six-gluon amplitudes in QCD}
\ShortTitle{Bootstrapping two-loop six-gluon amplitudes in QCD}

\author[a]{S\'ergio Carr\^olo}
\author[b]{Dmitry Chicherin}
\author*[a]{Johannes M. Henn}
\author[a]{Qinglin Yang}
\author[c,d,e]{Yang Zhang}

\affiliation[a]{Max-Planck-Institut f\"ur Physik, Werner-Heisenberg-Institut,\\
  Boltzmannstr.~8, 85748 Garching, Germany}
\affiliation[b]{LAPTh, Universit\'e Savoie Mont Blanc, CNRS,\\
  B.P.~110, F-74941 Annecy-le-Vieux, France}
\affiliation[c]{Interdisciplinary Center for Theoretical Study, University of Science and Technology of China,\\
  Hefei, Anhui 230026, China}
\affiliation[d]{Peng Huanwu Center for Fundamental Theory,\\
  Hefei, Anhui 230026, China}
\affiliation[e]{Center for High Energy Physics, Peking University,\\
  Beijing 100871, People's Republic of China}

\emailAdd{scarrolo@mpp.mpg.de}
\emailAdd{chicherin@lapth.cnrs.fr}
\emailAdd{henn@mpp.mpg.de}
\emailAdd{qlyang@mpp.mpg.de}
\emailAdd{yzhphy@ustc.edu.cn}

\abstract{%
The maximally transcendental, or ``most complicated'', terms of gauge-theory
scattering amplitudes have long been singled out, following Lipatov and
collaborators, as those parts of a QCD amplitude that most closely mirror
maximally supersymmetric Yang--Mills theory. We report on a programme that turns
this observation into a practical computational tool. We show that the rational
prefactors multiplying the highest-weight special functions of planar QCD
amplitudes are governed by four-dimensional leading singularities, which can be
classified and evaluated using on-shell diagrams. The resulting prefactors are
manifestly conformally invariant and admit compact spinor-helicity expressions
that hold for arbitrary multiplicity. Combining this input with the recently
established two-loop six-particle function space, we set up a symbol bootstrap
and determine, for the first time, the maximal-weight symbol of the planar
two-loop six-gluon amplitude in massless QCD, first for the
${-}{-}{+}{+}{+}{+}$ helicity configuration and subsequently for all MHV
configurations. The answer is fixed uniquely by physical consistency
conditions, requires a reduced alphabet of only $137$ symbol letters, and
yields as a byproduct previously unknown two-loop triple-collinear and
double-soft splitting functions. We summarise the method and the results, and
outline the directions they open up.}

\FullConference{Loops and Legs in Quantum Field Theory (LL2026)\\
12--17 April, 2026\\
Bayreuth, Germany\\}

\begin{document}
\maketitle

\section{Introduction}

Scattering amplitudes connect the Standard Model Lagrangian to collider
measurements and have repeatedly exposed hidden structures of quantum field
theory. The phenomenological demand for precision at the Large Hadron Collider
keeps pushing computations to higher loop order and multiplicity, while each new
frontier reveals an analytic structure that is far richer, and often far simpler,
than a brute-force calculation would suggest~\cite{Elvang:2013cua,Badger:2023eqz}.

For massless five-particle scattering, the past decade has seen dramatic
progress: the relevant two-loop Feynman integrals are known analytically and
complete two-loop amplitudes are available for a range of $2\to3$ processes of
phenomenological interest. The natural question is what happens at six particles. There the obstacles are of two qualitatively different
kinds. On the one hand, the special functions become genuinely more complicated:
the planar (leading-colour) two-loop six-particle function space consists of
iterated integrals in seven independent kinematic variables, with up to $167$
symbol letters at weight four~\cite{Abreu:2024fpe,Henn:2025xrc}. On the other
hand, and more subtly, the rational prefactors multiplying these functions
proliferate. Already at five gluons their explicit form is highly
non-trivial~\cite{DeLaurentis:2023fpc,DeLaurentis:2025dxw}, and a naive treatment
quickly becomes prohibitive at six points.

It is useful to recall the schematic structure of a perturbative QCD amplitude,
\begin{equation}
  \mathcal{A} \;=\; \sum_{i,j} c_{ij}\, R_i\, f_j\,,
  \label{eq:structure}
\end{equation}
with transcendental special functions $f_j$, rational (or algebraic) prefactors
$R_i$, and kinematic-independent rational numbers $c_{ij}$. Bootstrapping such an
amplitude requires detailed control of \emph{both} the function space and the
prefactors. The six-point function space was recently obtained by traditional
means~\cite{Abreu:2024fpe,Henn:2025xrc}; the prefactors are the missing
ingredient supplied here.

We are guided by the principle of maximal transcendentality. Kotikov, Lipatov and
collaborators observed that the maximally transcendental parts of certain QCD
quantities, such as the twist-two anomalous dimensions, coincide with those of
maximally supersymmetric Yang--Mills theory (sYM), since gluons dominate the BFKL
equation~\cite{Kotikov:2002ab,Kotikov:2004er}; see also~\cite{Broadhurst:1996ur}.
The correspondence is not exact for amplitudes---already the one-loop
${-}{+}{-}{+}$ configuration is richer than its sYM
counterpart~\cite{Gehrmann:2011aa,Dixon:2017nat}---but the maximally
transcendental part stays strikingly simple. We compute precisely these ``most
complicated terms'' for two-loop six-gluon amplitudes and argue that, in their
coefficients, they are the \emph{simplest} part of the amplitude. The material is
based on refs.~\cite{Carrolo:2025prl,Carrolo:2025long}.

\section{The strategy: hard functions, maximal weight, and on-shell diagrams}
\label{sec:strategy}

Our approach combines five ideas: we work with leading-colour six-gluon
amplitudes (because of the available function space); we study infrared-finite
hard functions; we project onto the maximally
transcendental, ``most complicated'' terms; this exposes a hidden conformal
symmetry and a direct connection to on-shell diagrams; these simplifications
make a symbol bootstrap feasible, with physical constraints fixing the answer
uniquely. We develop each in turn.

\subsection{Hard functions}
We compute helicity amplitudes in the leading-colour approximation, keeping the
leading term in the large-$N_c$ limit while retaining fermion loops by holding
the ratio $N_f/N_c$ fixed. Rather than the amplitude itself we study the
infrared-finite hard function obtained by subtracting the universal infrared
poles,
\begin{equation}
  \mathcal{H}^{(2)} \;=\; \lim_{\epsilon\to 0}
  \Big[ \mathcal{A}^{(2)} - I^{(1)}\,\mathcal{A}^{(1)} - I^{(2)}\,\mathcal{A}^{(0)} \Big]\,,
  \label{eq:hardfunction}
\end{equation}
where $\mathcal{A}^{(L)}(\epsilon)$ are the colour-stripped $L$-loop amplitudes
and $I^{(1)},I^{(2)}$ are the universal infrared subtraction
operators~\cite{Catani:1998bh,Aybat:2006wq}. Hard functions are less
sensitive to the spurious effects of dimensional regularisation than ``naive''
finite remainders; in five-particle scattering this was reflected in
cancellations within the relevant function
space~\cite{Chicherin:2020umh}. Crucially, the hard function can be analysed with
genuinely four-dimensional methods, for both its functions and its coefficients.

\subsection{The maximal-weight projection}
At maximal transcendental weight (weight four at two loops) the prefactors
simplify dramatically. The maximal-weight part of a Feynman integral is
conjecturally computed from its ``$d\log$'' integrand, the corresponding
prefactors---the leading singularities---being given by the maximal residues of
the integrand~\cite{ArkaniHamed:2010gh,Henn:2021bobadilla}. The same circle of
ideas, relating uniform transcendental weight to $d\log$ forms, has been
instrumental in understanding how to obtain canonical differential equations for
Feynman integrals~\cite{Henn:2013pwa}. The leading singularities can be
computed for a suitably defined infrared-finite part, namely the hard function.
They are determined by generalised unitarity cuts; the expectation that, at
maximal weight, these cuts can be evaluated in four dimensions was
argued for in ref.~\cite{Henn:2021bobadilla}. Our first target is therefore the
weight-four part of the planar QCD amplitudes.

\section{Prefactors from on-shell diagrams}
\label{sec:prefactors}

\begin{figure}[t]
  \centering
  \begin{tikzpicture}[scale=0.72,
      bv/.style={circle,draw=black,fill=black,inner sep=1.4pt},
      gv/.style={circle,draw=black,fill=gray!65,inner sep=1.4pt},
      ln/.style={thick}]
    \node[bv] (TL) at (0,1.4) {};
    \node[gv] (TM) at (1.4,1.4) {};
    \node[bv] (TR) at (2.8,1.4) {};
    \node[gv] (BL) at (0,0) {};
    \node[bv] (BM) at (1.4,0) {};
    \node[gv] (BR) at (2.8,0) {};
    \draw[ln] (TL)--(TM)--(TR);
    \draw[ln] (BL)--(BM)--(BR);
    \draw[ln] (TL)--(BL);
    \draw[ln] (TR)--(BR);
    \draw[ln] (TM)--(BM);
    \draw[ln] (TL)--(-0.85,2.25) node[above left] {$1^-$};
    \draw[ln] (BL)--(-0.85,-0.85) node[below left] {$6^+$};
    \draw[ln] (TR)--(2.45,2.4) node[above] {$2^-$};
    \draw[ln] (TR)--(3.25,2.4) node[above] {$3^+$};
    \draw[ln] (TR)--(3.85,1.4) node[right] {$4^+$};
    \draw[ln] (BR)--(3.7,-0.85) node[below right] {$5^+$};
  \end{tikzpicture}
  \caption{A representative two-loop on-shell diagram, of double-box topology,
  giving the non-trivial leading singularity $R_{5,6}$ for the ${-}{-}{+}{+}{+}{+}$
  helicity configuration (after ref.~\cite{Carrolo:2025prl}). Black and grey
  nodes denote the two types of trivalent on-shell vertex. All six leading
  singularities $R_{i,j}$ ($2<i<j\le 6$) arise from analogous diagrams; together
  with the Parke--Taylor term $R_1$ they are evaluated in four dimensions.}
  \label{fig:onshell}
\end{figure}
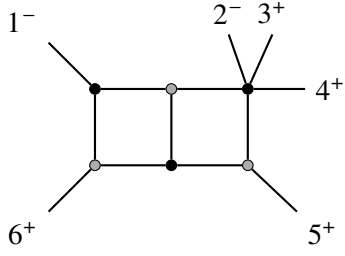

The decisive simplification concerns the prefactors multiplying the
maximal-weight functions. It is expected---and argued for in
ref.~\cite{Henn:2021bobadilla}---that these prefactors are four-dimensional
leading singularities. What can in addition be shown is that such leading
singularities are expressible in terms of on-shell diagrams. These diagrams are
built by gluing together three-point on-shell vertices, and the resulting
on-shell functions are easy to classify and to evaluate~\cite{ArkaniHamed:2012nw}.
A central observation is that this gluing procedure preserves conformal
invariance~\cite{Henn:2019mvc}: the leading singularities are annihilated by the
conformal generator
\begin{equation}
  k^{\alpha\dot\alpha} \;=\; \sum_{i=1}^{n}
  \frac{\partial}{\partial\lambda_{i\,\alpha}}\,
  \frac{\partial}{\partial\tilde\lambda_{i\,\dot\alpha}}\,,
  \label{eq:conformal}
\end{equation}
in line with the classical conformal invariance of massless
QCD~\cite{Witten:2003nn}. Here we use the spinor-helicity
formalism, in which massless momenta are written as
$p_i^{\alpha\dot\alpha}=\lambda_i^{\alpha}\tilde\lambda_i^{\dot\alpha}$ and the
holomorphic spinor brackets are denoted by
$\langle ij\rangle=\epsilon_{\alpha\beta}\lambda_i^{\alpha}\lambda_j^{\beta}$.
The conformal invariance is already manifest for the tree-level amplitude, which
for the ${-}{-}{+}\cdots{+}$ configuration is given by the Parke--Taylor factor
\begin{equation}
  \mathrm{PT}_{1,2} \;=\;
  \frac{\langle 12\rangle^4}{\langle 12\rangle\langle 23\rangle\langle 34\rangle
  \langle 45\rangle\langle 56\rangle\langle 61\rangle}\,.
  \label{eq:parketaylor}
\end{equation}
We focus on the ${-}{-}{+}{+}{+}{+}$ six-gluon configuration. At one loop, the
known amplitudes confirm that the coefficients of the highest-weight terms are
simply given by~\eqref{eq:parketaylor}, consistent with evaluating the relevant
on-shell box diagrams. At two loops, the Parke--Taylor prefactor
$R_1=\mathrm{PT}_{1,2}$ is supplemented by six new leading singularities $R_{i,j}$,
associated with the two-loop double-box topologies shown in
figure~\ref{fig:onshell}. The on-shell diagrams there represent maximal cuts of
these double boxes, on which all internal propagators are put on shell. A maximal
cut of the double box leaves a single residual integration, i.e.\ a one-form,
whose residue defines the leading singularity. Remarkably, a single closed
formula captures all of them,
\begin{equation}
  R_{i,j}/R_1 \;=\; -1 + 12\,u_{i,j} - 30\,u_{i,j}^2 + 20\,u_{i,j}^3\,,
  \qquad
  u_{i,j} := \frac{\langle 1i\rangle\langle 2j\rangle}{\langle 12\rangle\langle ij\rangle}\,.
  \label{eq:Rij}
\end{equation}
Since~\eqref{eq:Rij} is expressed in terms of holomorphic spinors only, its
conformal invariance is manifest. The same expression with $2<i<j\le n$
moreover captures all two-loop leading singularities of ${-}{-}{+}\cdots{+}$
amplitudes for arbitrary multiplicity $n$. As a consistency check, the known
maximal-weight coefficients of the five-particle two-loop
amplitudes~\cite{Abreu:2019odu,Agarwal:2023suw} reduce precisely to these
expressions. Although they look lengthy in the literature, at maximal weight they
collapse to the simple formula~\eqref{eq:Rij}, a simplification due in part to the
use of spinor-helicity variables.

\section{The symbol bootstrap}
\label{sec:bootstrap}

With the prefactors in hand, we can bootstrap the weight-four symbol of the
two-loop hard function. The \emph{symbol} of a transcendental function records
the sequence of differentials defining it as an iterated
integral~\cite{Goncharov:2010jf}, turning functional identities into linear
algebra. Symbol bootstraps have been powerful in planar $\mathcal{N}=4$
sYM~\cite{CaronHuot:2020bkp,Henn:2020omi}. Here we apply them for the first time to two-loop QCD amplitudes. We make an ansatz directly for the hard function of
eq.~\eqref{eq:hardfunction}, expanded on the seven leading singularities,
\begin{equation}
  \mathcal{H}^{(2)}_{\mathrm{YM}} \;=\; R_1\, G_1 \;+\!\!
  \sum_{2<i<j\le 6} R_{i,j}\, G_{i,j}\,,
  \label{eq:ansatz}
\end{equation}
where the $G_1, G_{i,j}$ are weight-four symbols from the space of planar two-loop
six-point functions, closed under cyclic
permutations~\cite{Abreu:2024fpe,Henn:2025xrc}. Only $G_1$ is affected by the
infrared subtraction (the tree-level and one-loop amplitudes involve only $R_1$),
which we implement by adding products of $I^{(L)}$ and lower-loop amplitudes. The
discrete flip symmetry $(123456)\leftrightarrow(216543)$ then leaves $2412$
unknowns.

The ansatz is then constrained by a sequence of physical consistency
requirements. First, the symbol entries of $G$ must be dimensionless. Second, the rational
prefactors $R_{i,j}$ contain spurious poles $1/\langle ij\rangle^3$ (for
non-adjacent $i,j$) and higher-order poles $1/\langle i\,i{+}1\rangle^4$ (for
adjacent momenta, understood cyclically), each of which must be cancelled by the
vanishing of the accompanying symbol at the corresponding locus. Third, the
amplitude must reproduce the known behaviour in the collinear and
triple-collinear limits, in which the hard function factorises into lower-point
hard functions and (infrared-subtracted) splitting functions, for all the
relevant helicity configurations. These conditions
reduce the initial $2412$ unknowns to a unique solution. Remarkably, they are
also mutually consistent---a non-trivial cross-check---and as a
bonus they determine the previously unknown triple-collinear and double-soft
splitting functions that enter them.

To reach full QCD we include fermion loops. In the planar limit the two-loop
hard function organises as
\begin{equation}
  \mathcal{H}^{(2)}_{\mathrm{QCD}} \;=\;
  \mathcal{H}^{(2)}_{\mathrm{YM}}
  + \frac{N_f}{N_c}\,\mathcal{H}^{[1]}
  + \Big(\frac{N_f}{N_c}\Big)^{2}\,\mathcal{H}^{[2]}\,.
  \label{eq:fermions}
\end{equation}
The fermionic sectors are simpler: the ${-}{-}{+}{+}{+}{+}$ configuration receives no
maximal-weight contribution from $\mathcal{H}^{[2]}$, while $\mathcal{H}^{[1]}$ is
controlled by six leading singularities of the same form,
\begin{equation}
  S_{i,j}/R_1 \;=\; -2 + 12\,u_{i,j} - 21\,u_{i,j}^2 + 11\,u_{i,j}^3\,,
  \label{eq:Sij}
\end{equation}
and is fixed by the first groups of conditions alone, without the
triple-collinear input.

\section{Novel results for scattering amplitudes and splitting functions}
\label{sec:results}

The bootstrap delivers the complete QCD answer for the maximal-weight symbol of
the planar two-loop ${-}{-}{+}{+}{+}{+}$ amplitude, and in subsequent
work~\cite{Carrolo:2025long} for all six-gluon MHV helicity configurations. Three
features deserve emphasis.

First, though the most complicated in terms of special functions, the
maximal-weight part is the simplest in its coefficients: its prefactors are the
few conformally invariant leading singularities of eqs.~\eqref{eq:Rij}
and~\eqref{eq:Sij}. That an answer consistent with all physical limits exists at
all is strong evidence for the result, and for the conjecture that these
prefactors are four-dimensional leading singularities.

Second, the symbol alphabet is reduced. Of the $245$ letters of the full two-loop
alphabet, $167$ can appear up to weight four, yet only $137$ occur in the final
answer. The same $137$-letter set governs all MHV configurations and has
independently appeared in studies of six-point Wilson loops with a Lagrangian
insertion. This reduction is at present unexplained and hints at deeper
structure---as in planar $\mathcal{N}=4$ sYM, where cluster adjacency and the
(extended) Steinmann relations proved instrumental in pushing the bootstrap to
 high loop orders~\cite{CaronHuot:2020bkp}---with first steps connecting some
of the six-particle letters to cluster-algebraic and flag-variety
descriptions~\cite{Bossinger:2025fes,Pokraka:2025nzh}. The QCD function space is,
however, far more complicated than its sYM counterpart---with many more letters
and less symmetry---so the analogy concerns structure rather than the size of the
problem.

Third, the calculation yields genuinely new physical data. Enforcing consistency
with the triple-collinear and double-soft limits determines the corresponding
two-loop splitting functions at maximal weight, extending results previously
known only at one loop: the maximal-weight double-soft function (for two gluons of the same helicity taken to be soft) coincides with
its $\mathcal{N}=4$ sYM counterpart, while the triple-collinear functions involve
a genuinely new weight-four ingredient. These feed directly into higher-point
amplitudes---the $g\to ggg$ symbol, for example, predicts part of the
leading-colour two-loop $gg\to Hgg$ amplitude---and are provided in the ancillary
files of refs.~\cite{Carrolo:2025prl,Carrolo:2025long}.

\section{Conclusions and outlook}
\label{sec:outlook}

The take-home message is twofold: the coefficients of the maximal-weight part of
QCD amplitudes are on-shell diagrams evaluable systematically in four dimensions;
and, with the symbol bootstrap, this fixes the planar six-gluon MHV amplitudes at
maximal weight---the ``most complicated'' terms proving the simplest to organise.

Several directions are now open.
\begin{itemize}
  \item \emph{From symbols to functions.} Repeating the bootstrap for iterated
  integrals rather than symbols poses no technical obstacle: it requires only the
  boundary behaviour at function level, which is standard, while the necessary
  expansions can conveniently be obtained via the defining differential equations
  of the functions, see e.g.~\cite{CaronHuot:2020grv}.
  \item \emph{Non-planar and NMHV.} The leading-singularity method extends
  naturally to non-planar contributions and, via prescriptive unitarity, to a
  reduced set of integrals and to the NMHV sector~\cite{Bourjaily:2017wjl}.
  \item \emph{Bypassing the integrals.} In the present work we relied on the
  explicitly known two-loop six-particle Feynman integrals. Combining the
  bootstrap with a Landau-equation analysis of the singularity structure may
  remove this prerequisite, eventually bypassing the explicit computation of
  Feynman integrals altogether.
  \item \emph{Below maximal weight.} The hardest question lies below maximal
  weight, where there are far fewer functions but many more coefficients:
  characterising the prefactors needed for QCD amplitudes---and how much
  four-dimensional information suffices---is already worth revisiting at one
  loop~\cite{Bern:1994zx,Dunbar:2009ax}.
  \item \emph{Approximate and numerical connections.} It would be interesting to
  connect this bootstrap philosophy to approximate numerical methods. If a
  candidate function reproduces the true amplitude in all relevant limits, by how
  much can it differ numerically elsewhere? One could envisage effective
  approximations in the spirit of the Pad\'e and conformal-mapping techniques
  developed for the Higgs-gluon form factor~\cite{Davies:2019nhm}, possibly
  exploiting recently discussed Stieltjes properties of perturbative
  expansions~\cite{Ditsch:2025stp}.
\end{itemize}

\acknowledgments
JMH and QY thank the participants of Loops and Legs 2026 for valuable
discussions. This work was supported by the European Union (ERC, UNIVERSE PLUS,
101118787); views and opinions expressed are those of the authors only. D.C.\ is
supported by ANR-24-CE31-7996 and Y.Z.\ by NSFC Grants No.~12575078 and
No.~12247103.

\end{document}